\documentclass[11pt,a4paper,fleqn]{article}

\usepackage[top=4cm, bottom=4cm, left=2cm, right=2cm]{geometry}
\usepackage{authblk}
\usepackage[utf8]{inputenc}
\usepackage{amsmath}
\usepackage{amsfonts}
\usepackage{amssymb}
\usepackage{amsthm}
\usepackage{hyperref}
\usepackage{tikz}
\usepackage{bbold}

\usepackage{bbm}

\usepackage[lined,ruled]{algorithm2e}
\SetKwRepeat{Do}{do}{while}

\usepackage{multirow}
\usepackage[lofdepth,lotdepth]{subfig}

\title{Biological and Shortest-Path Routing Procedures for Transportation Network Design}
\author{Fran\c cois Queyroi%
  \thanks{\texttt{francois.queyroi@parisgeo.cnrs.fr}}}
\affil{CNRS, UMR8504 Géographie-cités (CNRS/Université, Paris-1, Panthéon-Sorbonne/Université Paris Diderot)}

\date{}

\begin{document}
\flushbottom
\maketitle
\thispagestyle{empty}

\begin{abstract}
\textbf{Background.} The design of efficient transportation networks is an important challenge in many research areas. Among the most promising recent methods, biological routing mimic local rules found in nature. However  comparisons with other methods are rare. \\
\textbf{Methods.} In this paper we define a common framework to compare network design method. We use it to compare biological and a shortest-path routing approaches. \\
\textbf{Results.} We find that biological routing explore a more efficient set of solution when looking to design a network for uniformly distributed transfers. However, the difference between the two approaches is not as important for a skewed distribution of transfers.
\end{abstract}

\section{Introduction}

\textbf{Transportation networks.} The transportation network design  corresponds to the problem of selecting a set of possible links between locations (for example cities) in order for \emph{transfers} (for example of goods, people, \textit{etc.}) to be made possible~\cite{FRIESZ1985413}. The automated design of transportation network has a range of applications going from solving transshipment problems~\cite{gao2014biologically} to the computation of space trajectories~\cite{masi2014multidirectional}. In the social sciences, researchers want to compare efficient simulated networks with the real ones (railroads, railways, \textit{etc.}) in order to assess the existence and nature of suboptimal choices~\cite{mimeur:hal-01616746}. Transportation networks are also important for the simulation of the development of a city system~\cite{DBLP:journals/corr/RaimbaultBD16}.\\
The design of transportation networks as a computational problem is part of a large domain of inquiry which is known  as network flow problems~\cite{ahuja2014network}. Transportation network design can be here described as a multi-objective variant of the \emph{multi-commodity flow problem}~\cite{goldberg1998implementation} with unlimited edge capacities. \\

While most analytical research focus on cost minimisation, we are looking for transportation networks that are efficient with respect to multiple criteria. The choice therefore often involves a cost/benefits analysis. The most common criteria are the time performance (how quickly can we travel using the network), the cost (the size or total length of the network) and the tolerance to fault (how travel is affected by random perturbations). Obviously, finding a good balance between these criteria make the problem hardly solvable analytically and solutions are often explored using heuristics. \\

\textbf{Networks computation models.} Several computation models can be used to explore potential solutions. The first approach is what we call the \textit{biological} approach. The reason for this name is that those methods are derived from actual natural phenomena and the most cited one is probably the behaviour of the \textit{Physarum polycephalum } organism~\cite{vogel2016direct}. This slime mould development is indeed capable to solving mazes or discover shortest-paths in a difficult terrain~\cite{tero2007mathematical}. Experiments involving this organism caught a lot of attention in the scientific press. One important achievement was the simulation of the Greater Tokyo Area transportation network~\cite{tero2010rules} where the authors introduce an algorithm replicating the behaviour of the organism. It should be noted however that similar behaviours can be found in other natural phenomena  such as ant colonies~\cite{dussutour2004optimal} or current in an electrical network~\cite{ma2013current}. \\

The second possible approach to the problem of network design is what we call the \emph{shortest-path routing} method. We should stress out the fact that this category has known less extensions and led to less applications than the  previous one. To the best of our knowledge, Levinson and Yerra~\cite{levinson2006self} are the first to study this model of computation. Their objective was to show that a hierarchy of routes can emerge in the transportation network from a uniform distribution of transfers~\cite{yerra2005emergence}. This method has been rediscovered by others in the domain of Information Visualisation~\cite{lambert2010winding}.\\

Others methods such as greedy algorithms could be used. The idea is here to incrementally build a network by adding at each iteration the links that contribute the most to the performance of the network~\cite{demaine2010minimizing,parotisidis2015selecting}. A common variant is to start from a minimum cost spanning tree covering all transfers and then iteratively adding the best alternative routes. The \textit{Physarium} simulation was actually compared to this approach in \cite{tero2010rules}. It is also possible to start with a more complete network and prune the least used paths. This last method actually mimics the development of neural networks~\cite{navlakha2015decreasing}.\\

\textbf{Routing and Reinforcement.}  Both the biological and shortest-path approaches actually rely on two common mechanisms. First, the \textit{routing} of goods (food in the case of the slime mould) is done by assigning them to paths depending of their ``attractiveness'' (the diameter of the tubes for the slime mould). Then, paths attractiveness is updated according to the amount of goods using these paths (the slime mould's tubes expand or shrink due to the pressure). We call this second phase \textit{reinforcement}. By repeating those steps we can mimic a continuous process where the network gradually appears from a starting grid as least used paths in the grid gradually disappear while others gather more and more transfers.\\ 
Biological routing uses local rules of dispersion. Transfers will be assumed to behave like a liquid flowing through pipes of various size to reach a sink. On the other hand, the shortest-path method routes the transfers along the shortest-path between their source and destination. Biological routing therefore explores different paths even if there are longer while the shortest-path routing only selects the best paths.\\ 

\textbf{Contribution.}  The aim of this study is to analyse the differences between biological and shortest-path approaches. We introduce a common framework for the two algorithms. Most of the previous studies (in particular with the biological models) focus on uniform transfers between locations (either there is an exchange of commodities or not). However this setting may not be appropriate for different applications (in urban planning for example) since locations may differ in term of attraction potential. We therefore expand previous definitions of the algorithms in order to take into account arbitrary transfers distributions between several sources and destinations.\\
This common framework allows us to compare biological and shortest-path routing only focusing on the way transfers are routed without the interference of other minor differences. Moreover, we use as experiments a random generated set of grid-embedded locations and transfers while previous studies use a few small toy examples or real world configurations.\\

Biological network design such as the one simulating the \textit{Physarum polycephalum} organism has been shown to produce efficient network but was never, to the best of our knowledge, compared to the shortest-path approach. Our hypothesis is that the shortest-path method while not being as popular as biologically based approaches in the literature may be worthwhile to pursue if the transfers between location are modelled according to distributions found in human mobility (such as the gravity model).\\

\section{Reinforced Routing Procedure Overview}\label{sec:procedure}

\textbf{Notations.} Throughout this paper, we call the support graph $G=(V,E,l)$ a graph (or network) with nodes  (or vertices) set $V$, edge set $E$ and edge length $l : E \rightarrow \mathbb{R}^{+}$. Let $n=|V|$ the number of nodes and $m=|E|$ the number of edges. We call $F : V \times V \rightarrow \mathbb{R}^{+}$ the \textit{transfer matrix} between nodes in $G$ and we call $Q :  E \rightarrow \mathbb{R}^{+}$ a \textit{flow} distribution on the network (\textit{i.e.} the way transfers in $F$ are distributed along the edges of $G$). We have $\mathcal{F}=\sum_{s,t} F(s,t)$ the total amount of transfers between the nodes of $G$. A \textit{transportation network} on the support graph $G$ is simply a subgraph $N=(V,E',l)$ of $G$ with $E' \subseteq E$.
For edge defined functions such as $Q$ or $l$ and an edge $e=(u,v)$, we may write $Q(u,v)$ or $l(u,v)$ to refer to $Q(e)$ or $l(e)$. \\

Networks flows are mostly defined for transfers between a source node and one or several destination nodes. 
Here $Q$ corresponds to the distribution of the total $\mathcal{F}$ units among $E$ \textit{i.e.} if all units transferred travel through the network $G$ to reach their destination according to $F$ then $Q$ corresponds to the number of units that went through each edge. Notice that, for a given $G$ and $F$, there are multiple possible values of $Q$. Although we use the same terminology of flow, we do not expect $Q$ to respect to classic networks flow rules~\cite{ahuja2014network} and we consider $G$ to be undirected without a loss of generality. \\

\textbf{Main procedure backbone.} We design the procedure so that the differences between biological and shortest-path approaches only rest on the way transfers are routed.  As previously explained, the procedure can be divided into two parts: the routing of the transfers in $F$ (which gives $Q$) and the adaptation of edge length according to the flow $Q$. The routing depends on the length of the edges of the network $l$.  Algorithm \ref{alg:routing_procedure} details this generic procedure. Here, the reinforcement modifies the length of network edges by modifying $\sigma$ which can be interpreted as edges' ``diameter" of tubes in the slime mould organism,  the ``speed" in a railway network or the ``resistance'' in an electrical circuit. Notice that Algorithm \ref{alg:routing_procedure} does not directly produce a transportation network. Rather, the $\sigma$ values indicate whether an edge is likely to be part of an efficient transportation network. For the experiments, we create a network by selecting edges with a $\sigma$ value higher than $\epsilon$.\\

\begin{algorithm}
\LinesNumbered
\KwIn{$G=(V,E,l)$, $F : V \times V \rightarrow \mathbb{R}^{+}$, $\alpha>0$, $\mu > 1$, $\delta \in ]0,1]$, $\epsilon > 0$}
\KwOut{$\sigma : E \rightarrow [0,1]$}

$\sigma^0(e) \leftarrow 1$, $e \in E$\;
$n \leftarrow 0$\;

\Do{$\max |\sigma^{n+1} - \sigma^{n}|> \epsilon $}{
	$G' \leftarrow G \left( V,E,\dfrac{l}{\sigma^n} \right)$\;
	$Q \leftarrow$ \texttt{FlowRouting}$(G',F)$\;
	$\sigma^{n+1} \leftarrow \delta f(Q;\mu,\alpha) + (1-\delta)\sigma^{n}$\;
	$n \leftarrow n+1$\;
}
\Return $\sigma^{n}$\;
\caption{Reinforced Routing  of flow $F$ along network $G$}
\label{alg:routing_procedure}
\end{algorithm}

\textbf{Flow routing.} The function \texttt{FlowRouting} (line 5) will depend on the chosen method (Biological or Shortest-Paths routing). 
We assume that, in both cases, the result $Q$ is a normalized flow \textit{i.e.} such that $Q \in [0,1]^{m}$. To do this we simply divide the number of units transferred going through a given edge by $\mathcal{F}$. The two following sections will describe each routing procedure in details. 
\\

\textbf{Reinforcement.} First, notice $Q$ is an aggregation of transfers with different sources and destinations. Previous works~\cite{tero2010rules} used to adapt network edges after the routing of a single line of $F$  (the transfers connecting a single node $s$) or a single exchange in $F$ (the transfers between $s$ and $t$). This obviously requires to take lines or elements of $F$ in a random order. This approach may speed up the convergence of the procedure but we believe this addition of randomness is of little interest in the context of our study.  \\

In Algorithm \ref{alg:routing_procedure}, support network's edges' resistance $\sigma$ is modified according to a reinforcement function $f$ and an update rate $\delta$ (line 6). The latter is a parameter that influences the convergence rate of the algorithm. In previous studies, this parameter was often implicit: the reinforcement is defined as a change in resistance over time in a continuous fashion so the convergence speed is given by the way time is split into steps.\\ The reinforcement function (sometimes called ``value function'') $f$ we use is a logit-like function  frequently  used in the literature~\cite{tero2007mathematical}.

\begin{equation}
\label{eq:update_function} f(Q ; \mu,\alpha) = \dfrac{(1+\alpha)Q^{\mu}}{1+\alpha Q^{\mu}}
\end{equation}

\begin{figure}
\center
\includegraphics[scale=0.4]{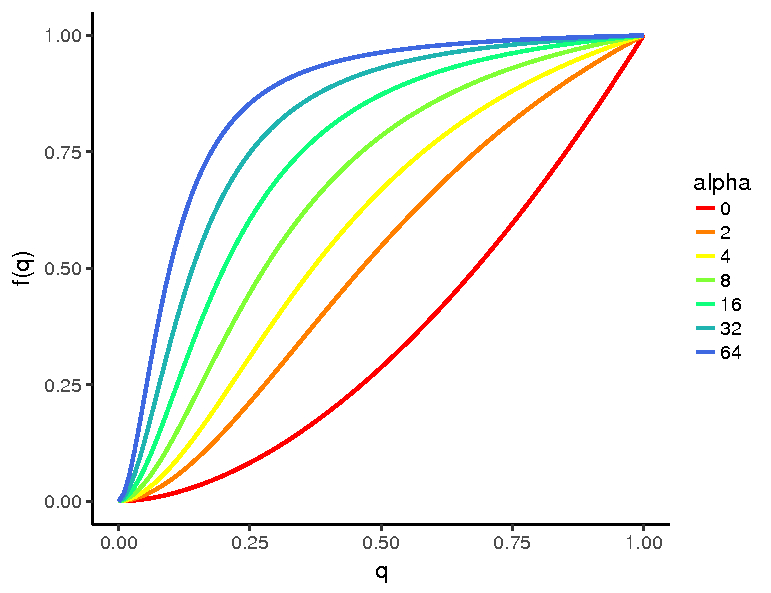}
\caption{\label{fig:fq_alpha_ex} $f(q;\mu,\alpha)$ for $\mu=1.8$ and different values of $\alpha$}
\end{figure}

Function $f$ is a response function where parameter $\mu>1$ controls the slope of the curve while parameter $\alpha>0$ controls the inflection point of the function. Indeed the higher $\alpha$ is the greater is the response value of small signals (see examples of curves with various values of $\alpha$ in Fig.~\ref{fig:fq_alpha_ex}). In the biology analogy, this function models the responses to the flow going through the Physarium tubes. We initialize $\sigma^{0}(e) = 1, \forall e \in E$ (line 1) and, since $f$ is dimensionless with $f(0)=0$ and $f(1)=1$, we have $\sigma^n \in [0,1]^{m}$ at each iteration of the algorithm. The length edges of the modified support graph $G'$ will increase (line 4) which will affect the next routing phase. \\

\textbf{Running time.} The stopping condition of the algorithm (line 8) depends of the difference in the distribution of $\sigma$ values between successive steps. The parameter $\epsilon$ is used to set the precision of the algorithm (in this study we used $\epsilon=5.10^{-4}$). Previous definition of the \textit{Physarium} algorithm uses a arbitrary number of iterations instead.\\
The convergence of resistance $\sigma$ to a stable solution  has been studied in the case of the \textit{Physarium} solver~\cite{miyaji2008physarum,ito2011convergence}. In our case, Algorithm~\ref{alg:routing_procedure} seems to always converge to a solution whether we use biological or shortest-path routing although the number of steps required for the shortest path routing procedure is lower.\\ 

The running time of function \texttt{FlowRouting} may also be affected by the ordering of the vertices. This is true whether we adopt the biological or the shortest-path routing procedure (described below). In both cases, we only need to route the transfers connected to the minimal vertex cover of the graph formed with the transfers in $F$ higher than $0$. The reason is that the flow we compute is an aggregation of different ``commodities'' (we can not exchange a person travelling from $a$ to $b$ with one going from $c$ to $d$). Both routing procedure will route one ``commodity'' (the people travelling from $a$ to $b$ or from $a$ to $c$) at a time. 
We can find a small enough vertex cover by taking vertices in decreasing order of their number of strictly positive transfers.\\

An implementation of the two procedures is available at \url{https://github.com/fqueyroi/tulip_plugins/tree/master/TransportationNetworks} as plugins for the Tulip software~\cite{auber:hal-01359308} \url{http://tulip.labri.fr}.\\

\subsection{Shortest-path routing}

Using the shortest-path model, the flow going through support network's edges is given by what we call the \textit{flow betweenness}. Informally, flow betweenness corresponds to the total number of travellers going through a given edge if the travellers choose the fastest (shortest) route. In the general case, we should therefore have:
\begin{equation}
\label{eq:flow_betweeness} Q(e) = \sum_{(s,t) \in V \times V} \dfrac{\pi_e(s,t)}{\pi(s,t)}F(s,t)
\end{equation}
where $\pi_e(s,t)$ is the number of shortest-paths from node $s$ to $t$ going through edge $e$ and $\pi(s,t)$ is the total number of shortest paths from node $s$ to $t$. In practice, if we use continuous edge length on a connected network, we can assume that $\pi_e(s,t) \in \{0,1\}$ and $\pi(s,t)=1$ for all pairs $(s,t)$. This means that there is only one shortest-path between each pair of nodes in $G$.\\

On way to compute Eq.~\ref{eq:flow_betweeness} is therefore to compute all pairwise shortest-paths between the locations evolved in the transfers using a shortest-path algorithm such as Dijkstra's algorithm~\cite{dijkstra1959note}.
However, notice that if $F(s,t)=1$ for all $(s,t) \in V \times V$ then Eq.~\ref{eq:flow_betweeness} corresponds to the \emph{betweenness centrality} measure~\cite{brandes2001faster}. This measure is well-known in network analysis as it allows to identify vital elements of the network (hubs). It is possible to generalize the algorithm proposed in \cite{brandes2001faster} in order to take account of transfer values different than $1$. The computation of \texttt{FlowRouting} therefore runs in $\mathcal{O}(nm + n\log m)$. In practice it involves computing as many shortest-path tree as the size of the minimal vertex cover of transfers higher than $0$ (which is at most $n$).\\

\subsection{Biological routing}
As previously explained, ``biological'' routing is designed to mimic several known physical phenomena that can be found in biological organisms such as the slime mould \textit{Physarum Polycephalum}. As such, the routing of flow in biological models corresponds to a flow $Q$ that respects classic flow constraints and minimizes the total energy of the model:
\begin{equation}
\label{eq:flow_minenergy} \xi_{\sigma}(Q) = \sum_{e \in E} \dfrac{l(e)}{\sigma(e)} Q(e)
\end{equation}
This biological routing of flow is also commonly introduced as a solution of a linear equation system:
\begin{equation}
\label{eq:ohm} Q(u,v) = \dfrac{l(u,v)}{\sigma(u,v)} (p(u) - p(v))
\end{equation}
Eq.~\ref{eq:ohm} corresponds to Ohm's law for electrical circuit where $\dfrac{l(u,v)}{\sigma(u,v)}$ is the \emph{conductance} of edge $(u,v)$. 
This routing is therefore also close to other network analysis measures such as Eigenvector centrality or PageRank~\cite{chung2010pagerank}. The \textit{potential} $p(u)$ of a node $u$ to attract flows depends on $u$'s neighbours' potential \textit{w. r. t.} the conductance (speed or diameter) of adjacent edges. However, in our case, we are not interested in the potential value of nodes but only by the flow going through each edge.\\

The computation of the flow $Q$ in the biological model corresponds to a solution of a linear equation system. 
This computation can be cumbersome due to the number of variables. 
We adopt the approximation algorithm described in \cite{kelner2013simple}. It starts with a suboptimal solution where the flows are routed on a \emph{low-stretch} spanning tree (distance preserving tree). Then, we randomly select a ``short-cut'' edge $(a,b)$  that does not belong to that tree and send a portion of the flow going from $a$ to $b$ in the tree through that short-cut. Those modifications are called ``cycle updates''. The number of cycle updates performed is set so that the distribution of flow is a $\epsilon$-approximation of the solution of the linear equation system. This approach leads to a near-linear time computation of $\mathcal{O}(m\log^2 n \log \log n \log(\epsilon^{-1}))$.
Note however that a solution $Q$ to Eq.~\ref{eq:flow_minenergy} can only be found if all transfers in $F$ have for source a single unique vertex (in order to respect flow constraints). Again, we need to find as many solutions the size of the minimal vertex cover of non-null transfers (which is at most $n$). The running time of the biological routing therefore highly depends on the density of transfers. 

\section{Experiment details}\label{sec:experiment}

We present in this section the choices made for comparing the two transportation network construction models described above.\\

\textbf{Fitness Measures.} We introduce here the fitness measures used. We use concepts found previously in the literature~\cite{tero2010rules} (the three dimensions: performance, fault-tolerance and cost). The differences with previous definitions of performance or fault-tolerance come from the fact that we consider arbitrary transfers distributions.\\
Even though Algorithm \ref{alg:routing_procedure} outputs a real vector of speed/diameter $\sigma$, we take as transportation network $N=(V,\{e \in E : \sigma(e)> \epsilon\},l)$ for simplicity sake. We call $\bar{d}_N$ the diameter of $N$ (length of the longest shortest-path) and $\mathcal{F}=\sum_{s,t} F(s,t)$ the total amount of transfers on the network. We define the following indicators:

\begin{enumerate}
\item \textbf{Performance} $P$ corresponds to the total time taken for units to go from their source to their destination in $N$.
\begin{equation}
\label{eq:perf} P(N,F)=1- \dfrac{1}{\mathcal{F}\bar{d}_N }\sum_{s,t} F(s,t) d_N(s,t)
\end{equation}
where  $d_N(s,t)$ is the distance between $s$ and $t$ in $N$ (\textit{i.e.} the sum of the length of the edges along the shortest-path from $s$ to $t$). Notice we have $P(N) \in [0,1]$ and we say the network $N$ has good performance when $P(N)$ is close to $1$. The main difference with previous definitions of performance is that the amount of transfers is taken into account. Indeed, the measure $P$ corresponds to the mean time taken by travellers to reach there destination while previous definition (using uniform transfers) correspond to the mean time of travel for each pair of locations.

\item \textbf{Fault Tolerance} $FT$ corresponds to the proportion of transfers still able to reach destination after the removal of a random edge in $N$.
\begin{equation}
FT(N,F) = 1-\dfrac{1}{|E(N)|} \sum_{e \in E(N) } \dfrac{1}{\mathcal{F}} \sum_{(s,t) \in V\times V}  F(s,t) \mathbb{1}_{P_{N \setminus e}(s,t) \neq \emptyset}
\end{equation}
where $\mathbb{1}_{P_{N \setminus e}\neq \emptyset}(s,t)$ is equal to $1$ if there is still a path between $s$ et $t$ in $N$ after the removal of edge $e$ or $0$ otherwise. Using this normalisation, we have $FT(N) \in [0,1]$ and we say that $N$ is highly tolerant to fault when $FT(N)$ is close to $1$.
Notice that $FT$ is a generalisation of the classic metric that just focused on whether the network is disconnected or not (a measure used in \cite{tero2010rules}). If an area is loosely connected to the network, it may not affect $FT$ much if the amount of transfers with this area is relatively low. 

\item \textbf{Cost} $C$ corresponds to the normalized sum of the length of edges in $N$
\begin{equation}
C(N) = \dfrac{\sum_{e \in E(N)} l(e)}{\sum_{e \in E(G)} l(e)}
\end{equation}
Using this normalisation, we have $C(N) \in [0,1]$ and we say that $N$ is a costly network when $C(N)$ is close to $1$.

\end{enumerate}

Obviously, a costly network is more likely to have a higher tolerance to fault and performance. Therefore it is interesting to look at the ratio $P/C$ and $FT/C$ for comparison purpose.\\

\textbf{Samples.} In order to compare the two algorithms, we  first use synthetic data generated randomly. We sample a set of $150$ points in the $[0,1]^2$ plain and connect the points using a standard Delaunay triangulation. This random grid corresponds to the possible adjacencies of the future network (the support graph $G$). We then select a subset of $8$ points that will correspond to the sources and destinations of transfers. The matrix $F$ is then generated using two different models:
\begin{enumerate}
\item Uniform distribution: set $F(s,t)=1$ for all $(s,t)$. It is the same as in \cite{tero2010rules}.
\item Gravity model: set $F(s,t)=\dfrac{P(s)P(t)}{d^{E}(s,t)^\gamma}$ where $P : V \rightarrow \mathcal{R}$ is the \textit{population} of the nodes and $d^{E}(s,t)$ is the euclidean distance between $s$ and $t$. Here we set $\gamma = 1.2$. This model is often used in urban geography to model human mobility~\cite{rodrigue2009geography}.  The population $P$ is generated using an Zipf exponential model \textit{i.e.} the population exponentially decreases with the rank of the points by a factor of $1.5$ (the ranks of the $8$ locations being chosen randomly). This model produces few important centres of attractions and isolated areas.
\end{enumerate}

The way transfers are distributed corresponds to two experimental settings. In addition, we develop a third and a forth using as locations cities of the French region \textit{Pays-de-la-Loire} (West of France). The resulting grid can be seen in Fig.~\ref{fig:pdl_unif_ex_routing} (red nodes represent important cities in the region). We also analyse the difference between a uniform and a gravity-like distribution of transfers. For the latter, we use as locations weight the actual population of the cities according to the 1999 national census.\\

\textbf{Parameters choices.} Algorithm \ref{alg:routing_procedure} has many different parameters. The most influential however is the parameter $\alpha$. For the experiments, we take for $\alpha$ values powers of $2$ (see the different behaviour of $f$ in Fig.~\ref{fig:fq_alpha_ex}). 
In \cite{tero2010rules}, the authors choose to modify the total amount of transfers (which here corresponds to $\mathcal{F}$) since their reinforcement function does not include a parameter similar to $\alpha$. The effect is however similar.  Indeed, the higher $\alpha$ is, the more the route that are less used are given a high weight. In \cite{tero2010rules}, the more transfers amount there is, the more those routes are likely to be used. The $\alpha$ value can be used to influence the final cost of the network. Small $\alpha$ values are likely to result in a tree-like organisation of the transportation network while higher $\alpha$ values are more likely to produce to a grid-like organisation~\cite{louf2013emergence}.\\
Regarding the other parameters we use an update rate $\gamma = 0.5$, a slope of reinforcement $\mu=1.8$ (similar to the one in Fig.~\ref{fig:fq_alpha_ex}) and a precision of $\epsilon=5.10^{-4}$ (\textit{i.e.} the algorithm will stop when the greatest difference in $\sigma$ values is smaller than $5.10^{-4}$).\\

\textbf{Hypothesis on the difference between the two approaches.} 
Previous studies show that the biological routing is efficient when compared for example to greedy approaches~\cite{tero2010rules}. The efficiency here involves finding a good compromise between performance and fault tolerance. These results where made using uniformly distributed transfers between locations. Using the same distribution, we expect shortest-path routing to perform worst than biological routing. The reason is that shortest-paths explore fewer solutions and the procedure could quickly fall  in a local minima. However, we could expect the  results to be different when modifying the distribution of transfers. A skewed distribution of transfers should favour the shortest-path routing approach since the direct routing of the most important transfers along the fastest route will have the most impact.\\

\section{Results}\label{sec:results}
In this section, we discuss the results of the various experiments using the indicators of performance ($P$), fault tolerance ($FT$) and cost ($C$). We first look  at the statistical results for the synthetic experiments then we discuss some qualitative aspect of the results in the case of the geographical grid. For clarity purpose, we use $SP$ and $BIO$ to refer to the shortest-path routing and biological routing respectively. The two types of transfers distribution are referred as $UM$ for the uniformly distributed transfers and $GM$ for the gravity model. \\

\begin{figure}
\center
\includegraphics[scale=0.83]{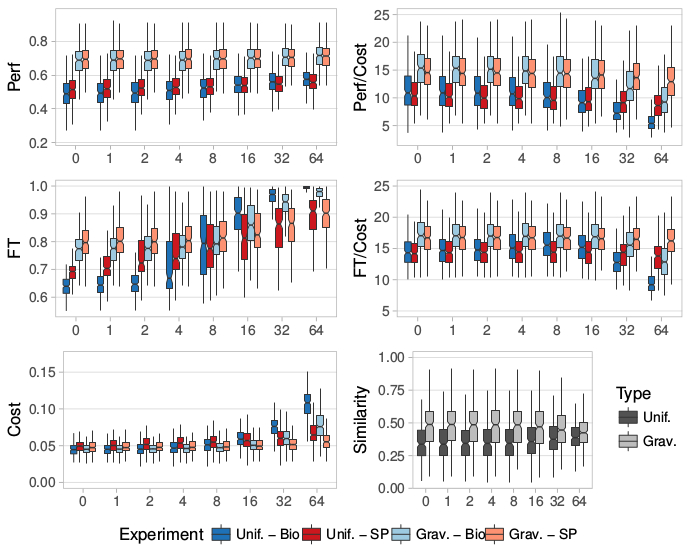}
\caption{\label{fig:barplot_alpha} Distribution of the various indicators according to $\alpha$ values (x-axis) for the four experimental settings (see Legend). Bottom-right plot: similarity between  the results of $SP$ and $BIO$ routing procedures (proportion of edges found in both networks).}
\end{figure}

\begin{figure}
\center
\includegraphics[scale=0.75]{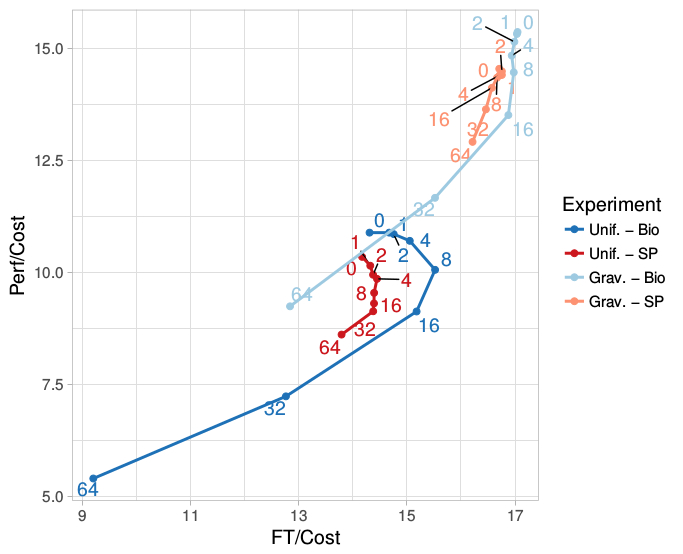}
\caption{\label{fig:evo_alpha_plot} Evolution of the median value of the ratio Perf/Cost and FT/Cost (middle of the corresponding boxplots in Fig.~\ref{fig:barplot_alpha}) according $\alpha$ values (points labels).}
\end{figure}

\begin{figure}[h]
  \begin{center}
    \subfloat[Bio routing -- $\alpha=4$]{
      \includegraphics[width=0.43\textwidth]{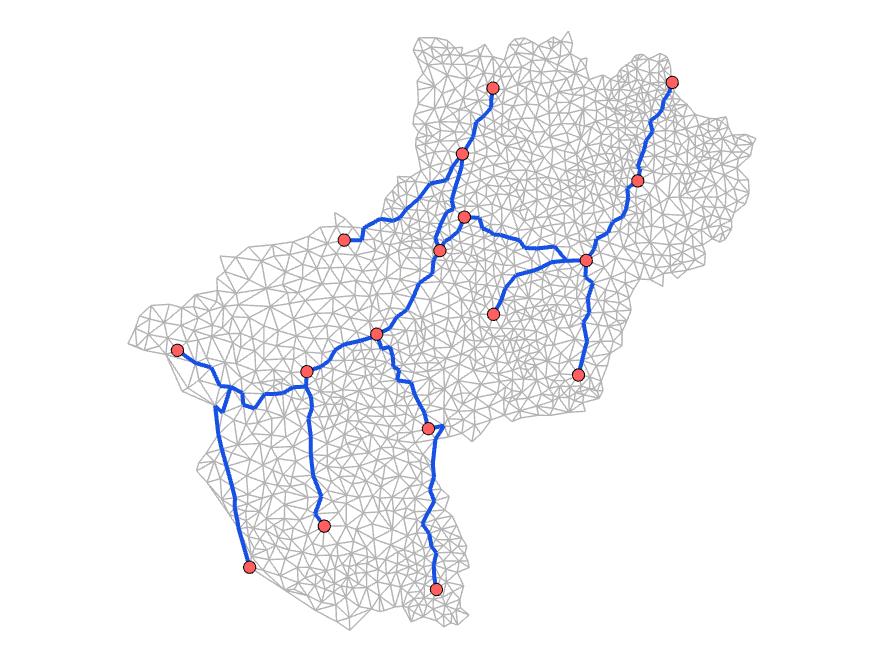}}
    \subfloat[Bio routing -- $\alpha=64$]{
      \includegraphics[width=0.43\textwidth]{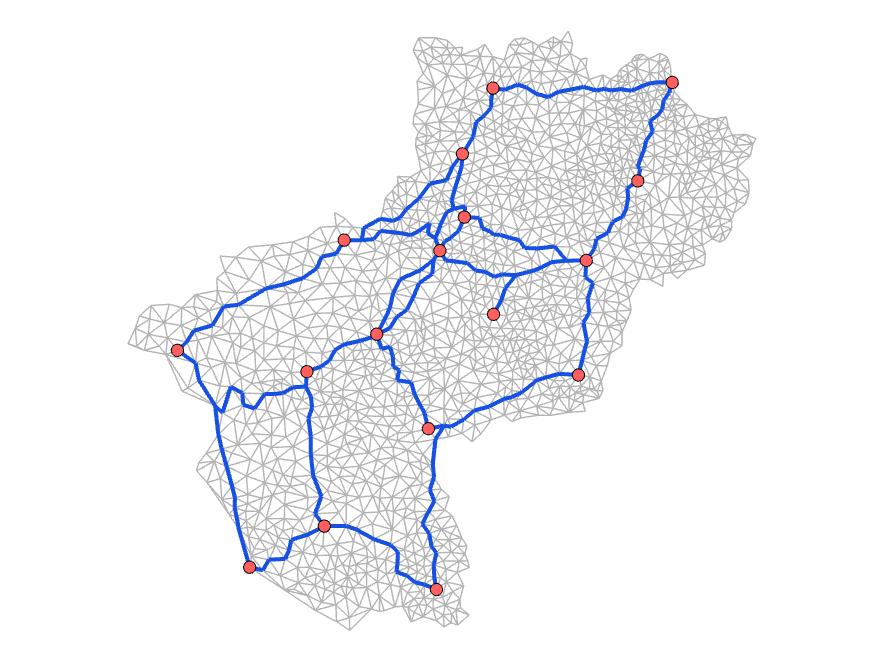}
      \label{sub:pdl_unif_ex_routing_bio64}}
        
     \subfloat[SP routing -- $\alpha=4$]{
      \includegraphics[width=0.43\textwidth]{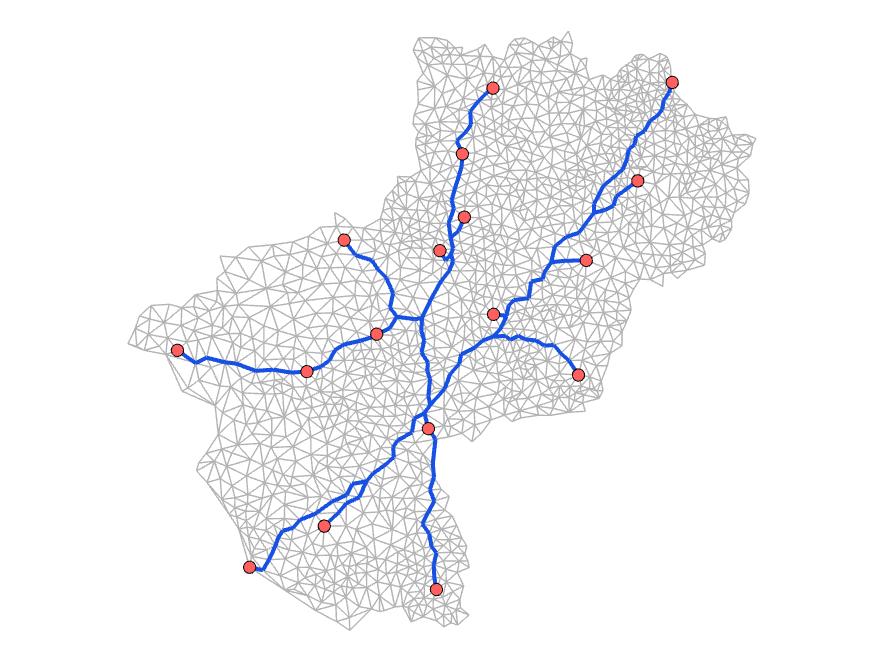}}    
      \subfloat[SP routing -- $\alpha=64$]{
      \includegraphics[width=0.43\textwidth]{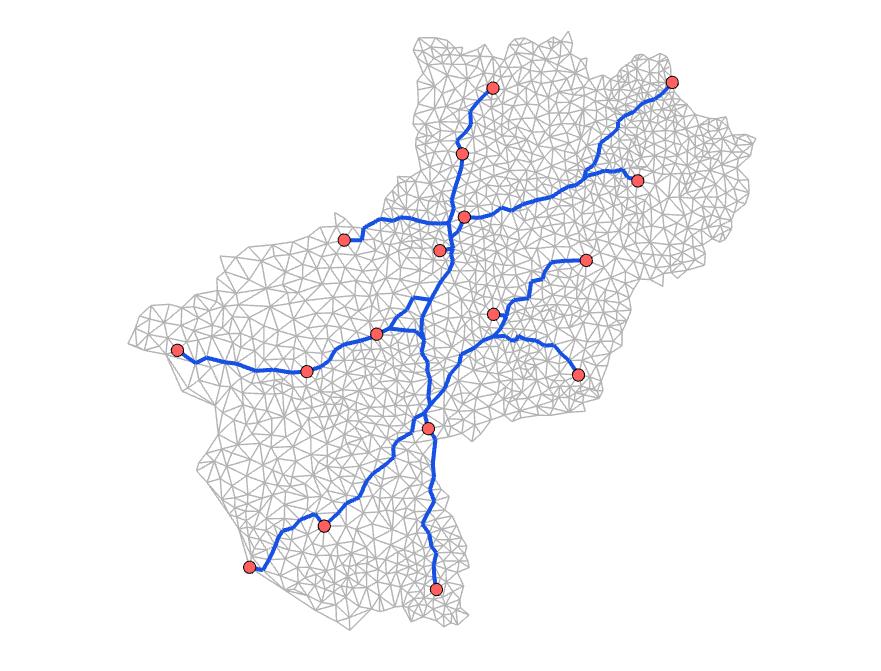}}
    \caption{Comparison of the result on the geographical example with uniform transfers. Red nodes correspond to major cities in the regions (\textit{préfecture} and \textit{sous-préfecture}).}
    \label{fig:pdl_unif_ex_routing}
  \end{center}
\end{figure}

\begin{figure}[h]
  \begin{center}
    \subfloat[Bio routing -- $\alpha=0$]{
      \includegraphics[width=0.43\textwidth]{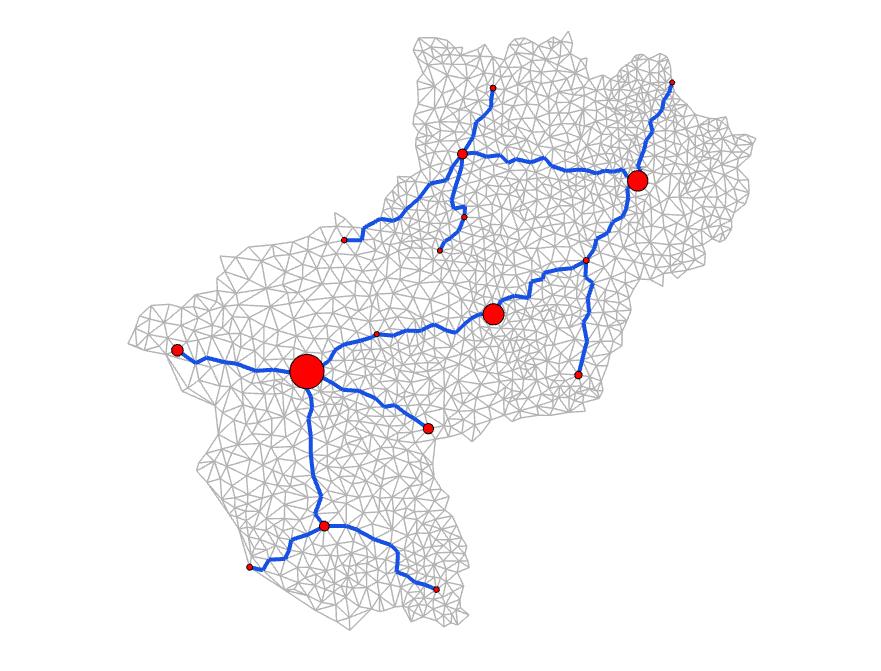}
      \label{sub:renonc}}
    \subfloat[Bio routing -- $\alpha=512$]{
      \includegraphics[width=0.43\textwidth]{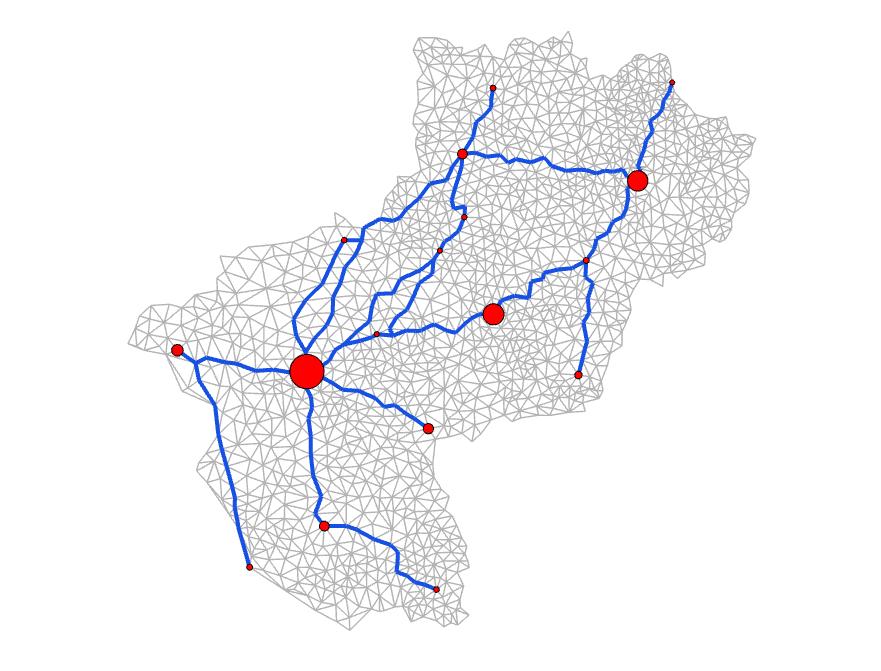}
      \label{sub:popul}}
        
     \subfloat[SP routing -- $\alpha=0$]{
      \includegraphics[width=0.43\textwidth]{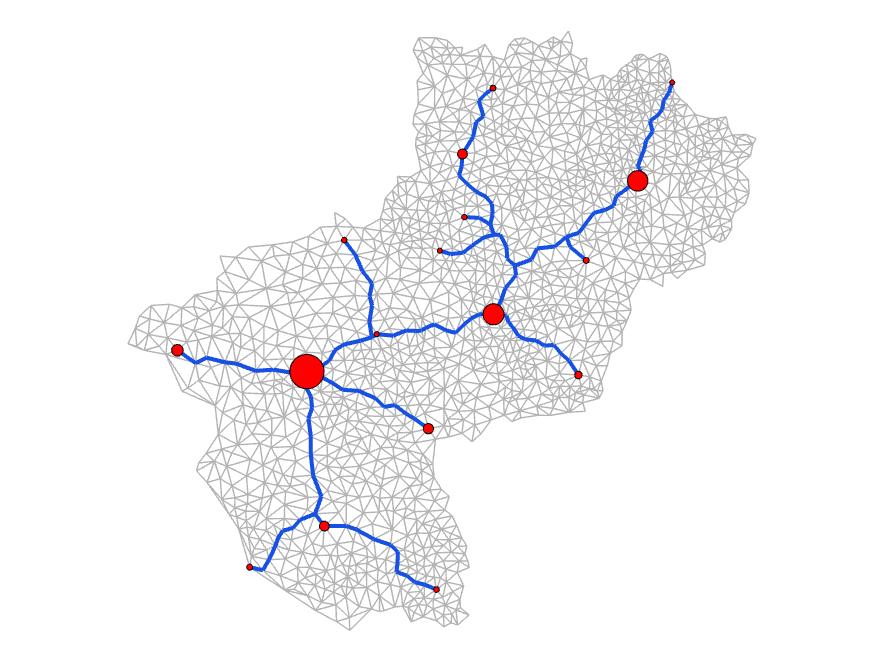}
      \label{sub:popul}}    
      \subfloat[SP routing -- $\alpha=512$]{
      \includegraphics[width=0.43\textwidth]{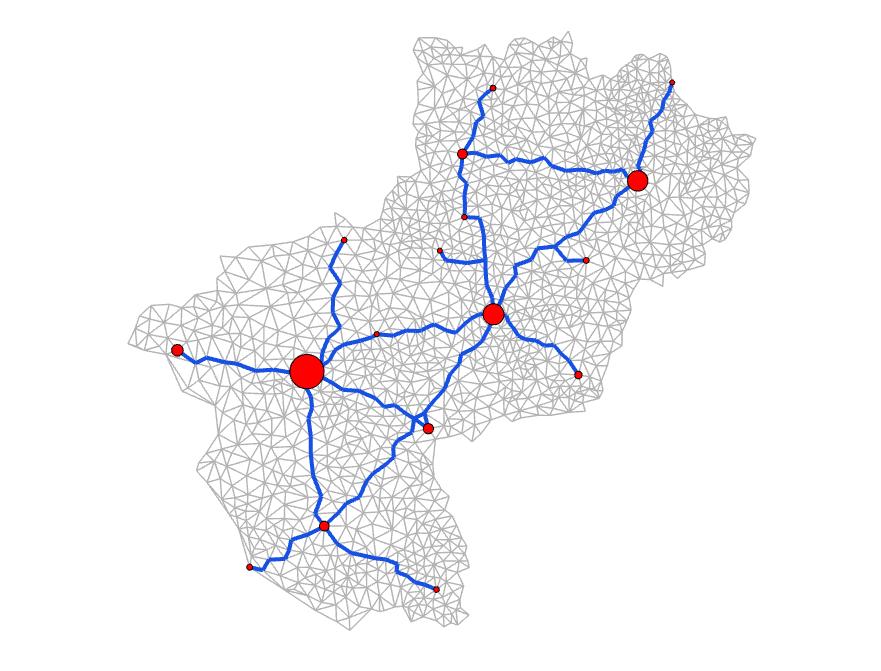}
      \label{sub:popul}}
    \caption{Comparison of the result on the geographical example with transfers following a gravity model. Red nodes correspond to major cities in the regions (\textit{préfecture} and \textit{sous-préfecture}). Node size is proportional to the population of the cities.}
    \label{fig:pdl_zipf_ex_routing}
  \end{center}
\end{figure}

We generated 200 instances of random grid and applied the transfers distribution models described above. We shall first stress out the fact that $BIO$ computation is very slow. It requires several minutes when $SP$ only takes a second.\\
The distributions of the indicators according to $\alpha$ can be found in Fig.~\ref{fig:barplot_alpha}. Note that we do not have to compare each algorithm using the same $\alpha$ value. Therefore, the evolution of the ratio $P/C$ and $FT/C$ can be found in Fig.~\ref{fig:evo_alpha_plot}. For the geographical grid, we report the statistics computed for some values of $\alpha$ for $UM$ (Table \ref{tab:geo_unif}) and $GM$ (Table \ref{tab:geo_zipf}). A representation of the computed networks can be seen in Figures \ref{fig:pdl_unif_ex_routing} and \ref{fig:pdl_zipf_ex_routing}. \\

Here are the conclusions that can be drawn from these results:
\begin{enumerate}
		\item \textbf{$BIO$ seems to achieve better results than $SP$ if we look at the ratio.}  In the $UM$ experiments and for $\alpha \in [1,4]$, the ratio $P/C$ and $FT/C$ are both greater for $BIO$ than for $SP$ with any $\alpha$ values. The same can be said in $GM$ with $\alpha <4$. However, the variation of the indicators is important (height of the boxplots in Fig.~\ref{fig:barplot_alpha}). It means there are configurations when $SP$ routing may find a better network. It is actually the case for the geographical grid with both $UM$ and $GM$. 
		
		\item \textbf{$BIO$ explores a broader set of results.} A trade-off between performance and fault-tolerance is clearly visible in $UM$ with both $BIO$ and $SP$. This phenomenon was already observed for $BIO$ in \cite{tero2010rules} using different fitness indicators.  One important difference between the two procedures is that the set of networks that can be found using $BIO$ corresponds to a wide range of cost. This is more limited for $SP$ as the networks found for various value of $\alpha$ may not differ a lot. 
		
		\item \textbf{There is important difference in the results between the two transfers distribution models}. The trade-off between perf6ormance and fault-tolerance is not clearly apparent in the $GM$ case. We can observe that the behaviour of the ratio $P/C$ and $FT/C$ is similar for $BIO$ and $SP$.  It can be explained by the fact that the simulated transfers create a single important centre of attraction in the grid (the most ``populated'' location). In this context, achieving good performance and fault-tolerance is not hard: we just need to connect this centre to the periphery. Accordingly, we observe that the networks found with $BIO$ and $SP$ are more similar in this model (see Fig.~\ref{fig:barplot_alpha}). Moreover, the value of the ratio $P/C$ and $FT/C$ is higher for both algorithms. Still, further expansions of the network (using higher $\alpha$ values) are less and less cost-effective since they connect locations whose transfers between them are exponentially smaller. 

		\item \textbf{The differences in performance or fault tolerance seems to correspond to different behaviours} when looking at the geographical example (Figs.~\ref{fig:pdl_unif_ex_routing} and \ref{fig:pdl_zipf_ex_routing}).\\ 
In the $UM$ case, networks found using $SP$ have a tree-like structure for most values of $\alpha$. $BIO$ finds similar networks using small $\alpha$ value. However, it can also find more costly networks with higher $\alpha$ value such as a grid-like network (see Fig.~\ref{sub:pdl_unif_ex_routing_bio64}). Table \ref{tab:geo_unif} shows this trade-off between $P/C$ and $FT/C$. Notice however that $BIO$ produces a lot of alternative paths that are actually parallel routes going between the same locations. This type of behaviour can explain the important increase of cost for higher $\alpha$ values.\\
In the $GM$ case, the result are very different. Small $\alpha$ values still correspond to tree-like structure for both $BIO$ and $SP$. For higher $\alpha$ values however, $SP$ produces alternative paths. This is also the case for $BIO$ but again with redundant routes. This apparently strange behaviour from $BIO$ could be traced back to the ``double edges'' often found in experimental settings using a real \textit{Physarum polycephalum} organism~\cite{miyaji2008failure}. In our case, it seems that the redundant edges do not add much in term of fault-tolerance since alternatives (but longer) routes already exist.
\end{enumerate}

\begin{table}[]
\centering
\caption{Results for the geographical grid (with a uniform distribution of transfers)}
\label{tab:geo_unif}
\begin{tabular}{l|l||lll|ll}
\hline
\textbf{Routing}                     & $\alpha$ & Perf & FT & Cost & Perf/Cost & FT/Cost \\ \hline
\multirow{3}{*}{SP}         &   2    &   .469     &  .793 & .034    &    13.794       &   23.324      \\ \cline{2-7} 
                            &   16    &   .479    &   .797  &  .035  &     13.686      &        22.771 \\ \cline{2-7} 
                            &   64    &   .489   & .822   &   .035   &     \textbf{13.971}      &        \textbf{23.486 }\\ \hline \hline 
\multirow{3}{*}{BIO} &   2    &   .471   &  .782  &  .035    &     \textbf{13.457}      &        22.343 \\ \cline{2-7} 
                            &   16    &  .487    &  .956  &  .042    &    11.595       &        \textbf{22.762} \\ \cline{2-7} 
                            &    64   &   .548   &  .998  &   .054   &     10.148      &        18.481 \\ \hline
\end{tabular}

\end{table}

\begin{table}[]
\centering
\caption{Results for the geographical network (with a gravity model distribution of transfers)}
\label{tab:geo_zipf}
\begin{tabular}{l|l||lll|ll}
\hline
\textbf{Routing}                     & $\alpha$ & Perf & FT & Cost & Perf/Cost & FT/Cost \\ \hline
\multirow{5}{*}{SP}         &  0 & .644 & .851 & .031 & \textbf{20.799} & \textbf{27.483}\\ \cline{2-7} 
                            &  128 & .661 & .928 & .034 & 19.369 & 27.201 \\ \cline{2-7} 
                            &  256 & .661 & .928 & .034 & 19.161 & 26.913\\ \cline{2-7} 
                            &  512 & .677 & .981 & .04 & 16.734 & 24.242 \\ \cline{2-7} 
                            &  1024 & .689 & .987 & .053 & 12.968 & 18.559 \\ \hline \hline

\multirow{5}{*}{BIO} &   0 & .578 & .831 & .033 & \textbf{17.343} & 24.914     \\ \cline{2-7} 
                            &   128 & .578 & .831 & .033 & \textbf{17.343} & 24.914      \\ \cline{2-7} 
                            &   256 & .627 & .938 & .037 & 16.743 &\textbf{ 25.05}    \\ \cline{2-7} 
                            &   512 & .637 & .967 & .045 & 14.15 & 21.469   \\ \cline{2-7} 

                            &   1024 & .689 & .987 & .053 & 10.759 & 15.398    \\ \hline
\end{tabular}

\end{table}

\section{Conclusion and Related Questions}\label{sec:conclusion}

In this paper we compared two different approaches of transportation network design. We provide a common analysis framework and an implementation of the algorithms. The methods were compared based on synthetic random grids using statistical indicators. From a quantitative point-of-view, we can conclude that the biological inspired approach is overall better than the routing of flows based on shortest-paths. However, the difference between the two is not always significant. The biologically-inspired model can be used to explore a wider range of solution. Contrary to our hypothesis, those observations are still valid when using a more realistic non-uniform model of transfers even though the difference in terms of performance and fault-tolerance is even smaller than in the uniform case.\\

Our experiments still have important limitations since there are numerous dimensions still unexplored. One could access, for example, the influence of the size of the grid. First results tend to reveal a similar behaviour but the discrepancy between the two methods may be amplified. The influence of others parameters such as the decay rate $\gamma$ or the reinforcement curve gradient $\mu$ should also be investigated but we expect they will have less of an impact.\\
Our study is also limited by the fact that our evaluation relies on objectives functions related to performance, fault-tolerance and cost. However, geographers, biologists or urban planners might want to know whether or not simulated networks are close to real-world networks. In this context, the objective functions are useful but it is not possible to infer the closeness between two networks based on the proximity in these functions values. \\

What we learn from this study is that biologically-inspired routing may be better suited for the researcher since it allows an exploration of a broader set of solutions. However, taking into account smaller and smaller flows (with increasing $\alpha$ value) leads to networks with high cost but without much additional performance or fault-tolerance.  One of our experiment reveals for example that biological routing may produce parallel short routes that are not efficient.

\bibliographystyle{plain}
\bibliography{bib_201707_multirouting}

\begin{thebibliography}{10}

\bibitem{ahuja2014network}
Ravindra~K Ahuja, Thomas~L Magnanti, and James~B Orlin.
\newblock {\em Network flows}.
\newblock Elsevier, 2014.

\bibitem{auber:hal-01359308}
David Auber, Romain Bourqui, Maylis Delest, Antoine Lambert, Patrick Mary, Guy
  Melan{\c c}on, Bruno Pinaud, Benjamin Renoust, and Jason Vallet.
\newblock {TULIP 4}.
\newblock Research report, {LaBRI - Laboratoire Bordelais de Recherche en
  Informatique}, September 2016.

\bibitem{brandes2001faster}
Ulrik Brandes.
\newblock A faster algorithm for betweenness centrality.
\newblock {\em Journal of Mathematical Sociology}, 25(2):163--177, 2001.

\bibitem{chung2010pagerank}
Fan Chung and Wenbo Zhao.
\newblock Pagerank and random walks on graphs.
\newblock In {\em Fete of combinatorics and computer science}, pages 43--62.
  Springer, 2010.

\bibitem{demaine2010minimizing}
Erik~D Demaine and Morteza Zadimoghaddam.
\newblock Minimizing the diameter of a network using shortcut edges.
\newblock In {\em Algorithm Theory-SWAT 2010}, pages 420--431. Springer, 2010.

\bibitem{dijkstra1959note}
Edsger~W Dijkstra.
\newblock A note on two problems in connexion with graphs.
\newblock {\em Numerische mathematik}, 1(1):269--271, 1959.

\bibitem{dussutour2004optimal}
Audrey Dussutour, Vincent Fourcassie, Dirk Helbing, and Jean-Louis Deneubourg.
\newblock Optimal traffic organization in ants under crowded conditions.
\newblock {\em Nature}, 428(6978):70, 2004.

\bibitem{FRIESZ1985413}
Terry~L. Friesz.
\newblock Transportation network equilibrium, design and aggregation: Key
  developments and research opportunities.
\newblock {\em Transportation Research Part A: General}, 19(5):413 -- 427,
  1985.
\newblock Special Issue Transportation Research: The State of the Art and
  Research Opportunities.

\bibitem{gao2014biologically}
Cai Gao, Chao Yan, Daijun Wei, Yong Hu, Sankaran Mahadevan, and Yong Deng.
\newblock A biologically inspired model for transshipment problem.
\newblock {\em arXiv preprint arXiv:1401.2181}, 2014.

\bibitem{goldberg1998implementation}
Andrew~V Goldberg, Jeffrey~D Oldham, Serge Plotkin, and Cliff Stein.
\newblock An implementation of a combinatorial approximation algorithm for
  minimum-cost multicommodity flow.
\newblock In {\em International Conference on Integer Programming and
  Combinatorial Optimization}, pages 338--352. Springer, 1998.

\bibitem{ito2011convergence}
Kentaro Ito, Anders Johansson, Toshiyuki Nakagaki, and Atsushi Tero.
\newblock Convergence properties for the physarum solver.
\newblock {\em arXiv preprint arXiv:1101.5249}, 2011.

\bibitem{kelner2013simple}
Jonathan~A Kelner, Lorenzo Orecchia, Aaron Sidford, and Zeyuan~Allen Zhu.
\newblock A simple, combinatorial algorithm for solving sdd systems in
  nearly-linear time.
\newblock In {\em Proceedings of the forty-fifth annual ACM symposium on Theory
  of computing}, pages 911--920. ACM, 2013.

\bibitem{lambert2010winding}
Antoine Lambert, Romain Bourqui, and David Auber.
\newblock Winding roads: Routing edges into bundles.
\newblock In {\em Computer Graphics Forum}, volume~29, pages 853--862. Wiley
  Online Library, 2010.

\bibitem{levinson2006self}
David Levinson and Bhanu Yerra.
\newblock Self-organization of surface transportation networks.
\newblock {\em Transportation Science}, 40(2):179--188, 2006.

\bibitem{louf2013emergence}
R{\'e}mi Louf, Pablo Jensen, and Marc Barthelemy.
\newblock Emergence of hierarchy in cost-driven growth of spatial networks.
\newblock {\em Proceedings of the National Academy of Sciences},
  110(22):8824--8829, 2013.

\bibitem{ma2013current}
Qi~Ma, Anders Johansson, Atsushi Tero, Toshiyuki Nakagaki, and David~JT
  Sumpter.
\newblock Current-reinforced random walks for constructing transport networks.
\newblock {\em Journal of the Royal Society Interface}, 10(80):20120864, 2013.

\bibitem{masi2014multidirectional}
Luca Masi and Massimiliano Vasile.
\newblock A multidirectional physarum solver for the automated design of space
  trajectories.
\newblock In {\em Evolutionary Computation (CEC), 2014 IEEE Congress on}, pages
  2992--2999. IEEE, 2014.

\bibitem{mimeur:hal-01616746}
Christophe Mimeur, Fran{\c c}ois Queyroi, Arnaud Banos, and Thomas
  Th{\'e}venin.
\newblock {Revisiting the structuring effect of transportation infrastructure:
  an empirical approach with the French Railway Network from 1860 to 1910}.
\newblock {\em {Historical Methods: A Journal of Quantitative and
  Interdisciplinary History}}, 2017.

\bibitem{miyaji2008physarum}
Tomoyuki Miyaji and Isamu Ohnishi.
\newblock Physarum can solve the shortest path problem on riemannian surface
  mathematically rigourously.
\newblock {\em International Journal of Pure and Applied Mathematics},
  47(3):353--369, 2008.

\bibitem{miyaji2008failure}
Tomoyuki Miyaji, Isamu Ohnishi, Atsushi Tero, and Toshiyuki Nakagaki.
\newblock Failure to the shortest path decision of an adaptive transport
  network with double edges in plasmodium system.
\newblock {\em International Journal of Dynamical Systems and Differential
  Equations}, 1(3):210--219, 2008.

\bibitem{navlakha2015decreasing}
Saket Navlakha, Alison~L Barth, and Ziv Bar-Joseph.
\newblock Decreasing-rate pruning optimizes the construction of efficient and
  robust distributed networks.
\newblock {\em PLoS computational biology}, 11(7):e1004347, 2015.

\bibitem{parotisidis2015selecting}
N~Parotisidis, Evaggelia Pitoura, and Panayiotis Tsaparas.
\newblock Selecting shortcuts for a smaller world.
\newblock In {\em SIAM International Conference on Data Mining (SDM)}, 2015.

\bibitem{DBLP:journals/corr/RaimbaultBD16}
Juste Raimbault, Arnaud Banos, and Ren{\'{e}} Doursat.
\newblock A hybrid network/grid model of urban morphogenesis and optimization.
\newblock {\em CoRR}, abs/1612.08552, 2016.

\bibitem{rodrigue2009geography}
Jean-Paul Rodrigue, Claude Comtois, and Brian Slack.
\newblock {\em The geography of transport systems}.
\newblock Routledge, 2009.

\bibitem{tero2007mathematical}
Atsushi Tero, Ryo Kobayashi, and Toshiyuki Nakagaki.
\newblock A mathematical model for adaptive transport network in path finding
  by true slime mold.
\newblock {\em Journal of theoretical biology}, 244(4):553--564, 2007.

\bibitem{tero2010rules}
Atsushi Tero, Seiji Takagi, Tetsu Saigusa, Kentaro Ito, Dan~P Bebber, Mark~D
  Fricker, Kenji Yumiki, Ryo Kobayashi, and Toshiyuki Nakagaki.
\newblock Rules for biologically inspired adaptive network design.
\newblock {\em Science}, 327(5964):439--442, 2010.

\bibitem{vogel2016direct}
David Vogel and Audrey Dussutour.
\newblock Direct transfer of learned behaviour via cell fusion in non-neural
  organisms.
\newblock In {\em Proc. R. Soc. B}, volume 283, page 20162382. The Royal
  Society, 2016.

\bibitem{yerra2005emergence}
Bhanu~M Yerra and David~M Levinson.
\newblock The emergence of hierarchy in transportation networks.
\newblock {\em The Annals of Regional Science}, 39(3):541--553, 2005.

\end{thebibliography}

%
%
%
%

\end{document}